\documentclass[floatfix,showpacs,pra,aps,twocolumn]{revtex4}
 \usepackage{bm}
 \usepackage{amsmath}
 \usepackage{amssymb}
 \usepackage{latexsym} 
 \usepackage{amsfonts} 
 \usepackage{epsfig}
 \usepackage{psfrag}

 \newcommand{\f}[1]{\mbox{\boldmath$#1$}}

 \newcommand{\ket}[1]{|#1\rangle} 
 \newcommand{\bra}[1]{\langle#1|} 
 \newcommand{\bracket}[2]
 {\langle#1|#2\rangle} 
 \newcommand{\nn}{\nonumber\\} 
 \newcommand{\bea}{\begin{eqnarray}}
 \newcommand{\ea}{\end{eqnarray}}
 \newcommand{\eea}{\end{eqnarray}}
 
 \newcommand{\ord}{{\cal O}}

\newcommand{\abs}[1]{\left|#1\right|} 
\newcommand{\pdiff}[2]{\frac{\partial#1}{\partial#2}} 
\newcommand{\tdiff}[2]{\frac{d#1}{d#2}} 

\begin{document}

\title{General error estimate for adiabatic quantum computing}

\author{Gernot Schaller, Sarah Mostame, and Ralf Sch\"utzhold$^*$}

\affiliation{Institut f\"ur Theoretische Physik, 
Technische Universit\"at Dresden, 01062 Dresden, Germany}

$^*$ email: {\tt schuetz@theory.phy.tu-dresden.de}

\begin{abstract} 
Most investigations devoted to the conditions for adiabatic quantum
computing are based on the first-order correction 
${\bra{\Psi_{\rm ground}(t)}\dot H(t)\ket{\Psi_{\rm excited}(t)}
/\Delta E^2(t)\ll1}$. 
However, it is demonstrated that this first-order correction does not 
yield a good estimate for the computational error.
Therefore, a more general criterion is proposed, which includes
higher-order corrections as well and shows that the computational
error can be made exponentially small 
-- which facilitates significantly shorter evolution times than the
above first-order estimate in certain situations. 
Based on this criterion and rather general arguments 
and assumptions,
it can be demonstrated that 
a run-time~$T$ of order of the inverse minimum energy 
gap~$\Delta E_{\rm min}$ is sufficient and necessary, i.e., 
$T=\ord(\Delta E_{\rm min}^{-1})$.
For some examples, these analytical investigations are confirmed by
numerical simulations. 
\end{abstract} 


\maketitle

\section{Introduction}
%
With the emergence of the first quantum algorithms, it turned out that 
quantum computers are in principle much better suited to solving
certain classes of problems than classical computers. 
Prominent examples are Shor's algorithm \cite{shor1997} for the
factorization of large numbers into their prime factors in polynomial
time and Grover's algorithm \cite{grover1997} for searching an unsorted
database with $N$ items reducing the computational complexity from the
classical value $\ord(N)$ to $\ord(\sqrt{N})$ on a quantum computer.   

Unfortunately, the actual realization of usual sequential quantum
algorithms (where a sequence of quantum gates is applied to some
initial quantum state, see, e.g., \cite{nielsen2000}) goes along with
the problem that errors accumulate over many operations and the
resulting decoherence tends to destroy the fragile quantum features
needed for the computation.
Therefore, an alternative scheme has been suggested
\cite{farhi2000}, where the solution to a problem is encoded in the
(unknown) ground state of a (known) Hamiltonian.
By starting with an initial Hamiltonian $H_{\rm i}$ with a known
ground state and slowly evolving to the final Hamiltonian 
$H_{\rm f}$ with the unknown ground state, e.g., 
$H(t)=[1-s(t)] H_{\rm i} + s(t) H_{\rm f}$, adiabatic quantum
computing makes use of the adiabatic theorem which states that a
system will remain near its ground state if the evolution $s(t)$ is
slow enough.
Since there is evidence that the ground state is more robust against
decoherence \cite{childs2001,kaminsky0211152,sarandy2005b}, this scheme 
offers fundamental advantages compared to sequential quantum algorithms.

However, determining the achievable speed-up of adiabatic quantum
algorithms (compared to classical methods) for many problems is still
a matter of investigation and debate, see, e.g., \cite{znidaric2005,
farhi0512159,aharonov0405098,sarandy2004,childs2002,das2003,roland2002}. 
For example, it has been argued in \cite{aharonov0405098} that all
conventional (sequential) quantum algorithms can be realized as
adiabatic quantum computation schemes with polynomial overhead via the 
history interpolation (polynomial equivalence).
For an adiabatic version of Grover's algorithm, a constant velocity
$\dot s$ implies a linear scaling of the run-time $T=\ord(N)$,  
whereas a suitably adapted time-dependence $s(t)$ yields the known 
quadratic speed-up $T=\ord(\sqrt{N})$, cf.~\cite{das2003,roland2002}.
Whether adiabatic algorithms of NP complete problems such as 3-SAT can
be even more efficient than this quadratic speed-up is still not
clear, see, e.g., \cite{znidaric2005,farhi0512159}.  

In this paper, we derive a general error estimate as a function of the
run-time $T$ (the main measure for the computational complexity
of adiabatic quantum algorithms) for very general gap structures  
$\Delta E(s)$ and interpolation velocities $s(t)$.  

\section{Adiabatic Expansion}
%
The evolution of a system state $\ket{\Psi(t)}$ subject to a
time-dependent Hamiltonian $H(t)$ is described by the
Schr\"odinger equation ($\hbar=1$)
\bea
\label{Eschroedinger}
i \ket{\dot{\Psi}(t)} = H(t) \ket{\Psi(t)}
\,.
\eea
Using the instantaneous energy eigenbasis defined by
$H(t) \ket{n(t)} = E_n(t) \ket{n(t)}$,
the system state $\ket{\Psi(t)}$ can be expanded to yield
\bea
\ket{\Psi(t)} = \sum_n a_n(t) \exp\left\{-i \int\limits_0^t E_n(t')
dt'\right\} \ket{n(t)}
\,.
\eea
Insertion into the Schr\"odinger equation yields -- after some algebra
-- the evolution equations for the coefficients 
\bea
\label{Ecoeff2}
\pdiff{}{t} \left(a_m e^{-i\gamma_m}\right)
&=&
-\sum_{n\neq m} a_n 
\,\frac{\bra{m}\dot{H}\ket{n}}{\Delta E_{nm}} 
\,e^{-i\gamma_m}
\times
\nn
&&
\times
\exp\left\{-i \int\limits_0^t\Delta E_{nm}(t')dt'\right\}
\eea
with the energy gap ${\Delta E_{nm}(t)=E_n(t)-E_m(t)}$ 
and the Berry phase \cite{sun1988}
\bea
\label{Eberryphase}
\gamma_n(t) = i \int\limits_0^t  dt'\,\bracket{n(t')}{\dot{n}(t')}
\,.
\eea
If the external time-dependence $\dot H$ is slow (adiabatic evolution),
the right-hand side of Eq.~(\ref{Ecoeff2}) is small and the solution
can be obtained perturbatively.
After an integration by parts, the first-order contribution yields
\bea
\label{Efirst_order}
a_m(t)
&\approx&
a_m^0e^{i\gamma_m(t)}
-i
\left[
\sum_{n\neq m}
a_n^0
\frac{\bra{m}\dot{H}\ket{n}}{\Delta E_{nm}^2}
\,e^{i\varphi_{nm}}
\right]_0^t
\eea
where $\varphi_{nm}\in\mathbb R$ denotes a pure phase.
Consequently, if the local adiabatic condition
\bea
\label{Eadiabatic_old}
\frac{\bra{m}\dot{H}\ket{n}}{\Delta E_{nm}^2} = \varepsilon \ll 1
\eea
is fulfilled for all times, the system approximately stays in its
instantaneous eigen (e.g., ground) state throughout the (adiabatic)
evolution. 
This above constraint has frequently been used as a condition for
adiabatic quantum computation \cite{farhi2000,childs2002}.
However, since the solution to a problem is encoded in the ground
state of the final Hamiltonian in adiabatic quantum computation
schemes, it is not really necessary to be in the instantaneous ground
state {\em during} the dynamics -- the essential point is to obtain
the desired ground state {\em after} the evolution. 
Since the external time-dependence $\dot H$ could realistically be 
extremely small (or even practically vanish) at the end of the
computation $t=T$, the first-order result (\ref{Efirst_order}) does
not always provide a good error estimate. 
Similar to the theory of quantum fields in curved space-times \cite{birrell},
the difference between the adiabatic and the instantaneous vacuum
should not be confused with real excitations (particle creation).
Therefore, it is necessary to go beyond the first-order result above and 
to estimate the higher-order contributions.

\section{Analytic Continuation}
%
Evidently, the Schr\"odinger equation is covariant under simultaneous
transformations of time and energy, such that the runtime of any 
adiabatic algorithm can be reduced to constant if the energy of the
system is modified accordingly \cite{das2003}. 
Here we want to exclude a mixing of these effects and will therefore 
assume
\bea
\label{Etraceconst}
{\rm Tr}\{H[s(t)]\} = {\rm const.} \qquad \forall\; s\in[0,1]
\,,
\eea
where $0 \le s(t) \le 1$ is an interpolation function which will be 
specified below.
In practice, the above condition can even be relaxed to the demand
that the trace should not vary by orders of magnitude
(during $0 \le s \le 1$).
With suitable initial and final Hamiltonians $H_{\rm i}$ and $H_{\rm f}$,
the above condition can be satisfied for all $s$ by using the linear
interpolation scheme
\bea
\label{Ehamiltonevolution}
H(t) = \left[1-s(t)\right] H_{\rm i} + s(t) H_{\rm f}\,,
\eea
but other schemes are also possible (see section \ref{Sextension}).
For simplicity, we restrict our considerations in this section to a
non-degenerate (instantaneous) ground state $n=0$ and one single first 
exited state $m=1$ with $\Delta E=\Delta E_{10}$.
(Multiple excited states will be discussed in section \ref{Sextension}.)
Similarly, all energies will be normalized in units of a typical
energy scale corresponding to the initial/final gap, i.e.,
$\Delta E(0)=\ord(1)$ and $\Delta E(1)=\ord(1)$.
We classify the dynamics of $s(t)$ via a function 
$h(s)\geq 0$
\bea
\label{Eclassify}
\tdiff{s}{t} = \Delta E(s) h(s)
\,,
\eea
where the function $h(s)\geq0$ is constrained by the conditions 
$s(0)=0$ and $s(T)=1$. 
Insertion of this ansatz into Eq.~(\ref{Ecoeff2}) yields the 
exact formal expression for the non-adiabatic corrections to a system
starting in the ground state, i.~e., with $a_1(0)=0$ one obtains after
time $T$ 
\bea
\label{Eformal}
a_1(1) e^{-i \gamma_1(1)} 
&=& 
-  
\int\limits_0^1
ds\;
a_0(s)e^{-i \gamma_1(s)} 
\,\frac{F_{01}(s)}{\Delta E(s)}
\times 
\nn
&&
\times 
\exp\left\{-i\int\limits_0^{s}\frac{ds'}{h(s')}\right\}
\,,
\eea
with the matrix elements $F_{nm}(s)=\bra{m(s)}H'(s)\ket{n(s)}$ which
simplify in the case~(\ref{Ehamiltonevolution}) of linear
interpolation to 
$F_{nm}(s)=\bra{m(s)}(H_{\rm f}-H_{\rm i})\ket{n(s)}$. 
The advantage of the form in Eqs.~(\ref{Eclassify}) and
(\ref{Eformal}) lies in the fact that different time-dependences
$s(t)$ and hence different choices for $h(s)$ solely modify the
exponent.  

\begin{figure}
\includegraphics[height=3cm]{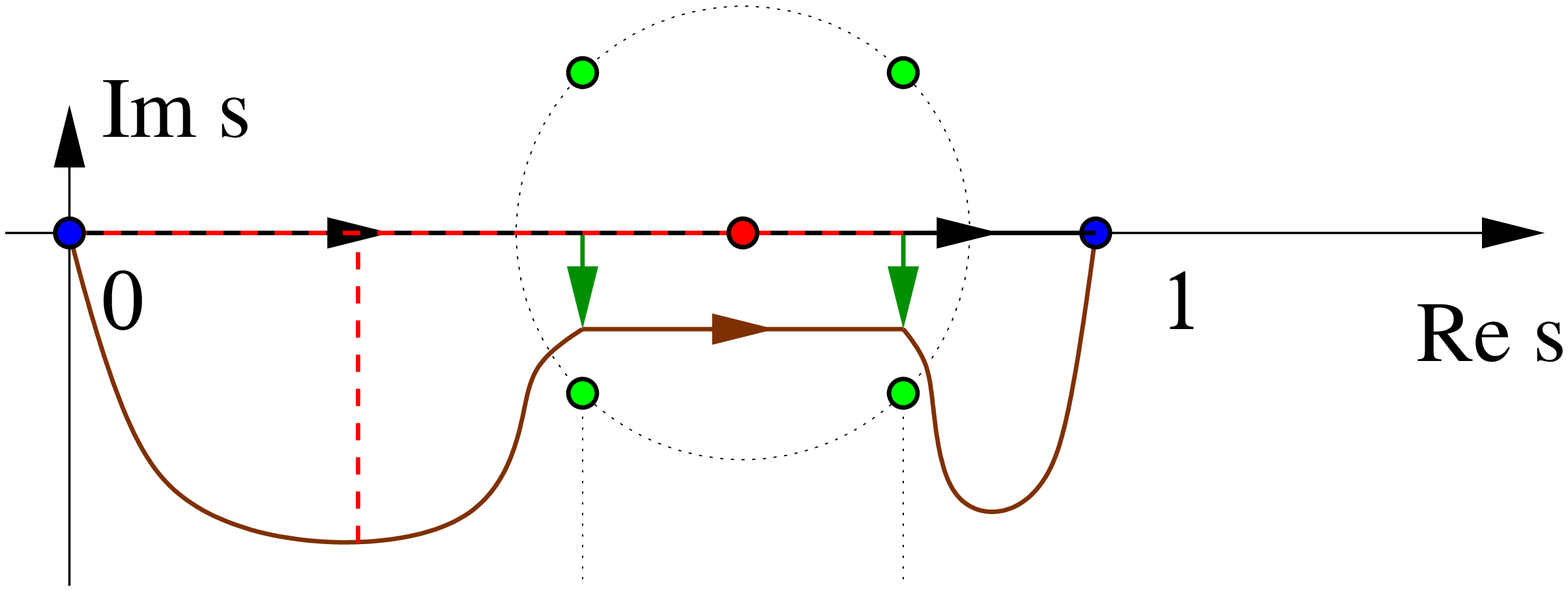}
\caption{\label{Fintpath}[Color Online] 
The original integration contour (black line along real axis) of
equation~(\ref{Eformal}) is shifted to the complex plane 
(curved line).  
The gap structure $\Delta E(s)$ leads to singularities near the real
axis [green hollow circles, here displayed for $2a=4$ in
  Eq.~(\ref{Eenergy_gap})],  
which limit the deformation of the integration contour.
The integral in the exponent (dashed line) in equation~(\ref{Eformal}) 
ranges from $0$ to $s'$, which gives rise to a real contribution to
the exponent off the real axis only.}
\end{figure}

We assume that all involved functions can be analytically continued 
into the complex $s$-plane and are well-behaved near the real
$s$-axis. 
Given this assumption, we may estimate the integral in
Eq.~(\ref{Eformal}) via deforming the integration contour into the
lower complex half-plane (to obtain a negative exponent -- which is
the usual procedure in such estimates) until we hit a saddle point, a
singularity, or a branch cut, see Fig.~\ref{Fintpath}.  
Deforming the integration contour into the upper complex half-plane
would of course not change the result, but there the integrand is
exponentially large and strongly oscillating such that the integral is
hard to estimate.
Since the gap $\Delta E(s)$ usually has a pronounced minimum at 
$s_{\rm min}\in(0,1)$, the first obstacle we encounter
\cite{well-behaved} 
will be a singularity at $\tilde{s}$ close to the real axis, i.e.,
$\abs{\Im(\tilde{s})}\ll1$ and $\Re(\tilde{s})\approx s_{\rm min}$, 
where $\Delta E(\tilde{s})=0$.

Let us first consider a constant function $h(s)=h$:
Assuming $h \ll 1$ (i.e., slow evolution), the exponent in
Eq.~(\ref{Eformal}) acquires a large negative real part for $\Im(s)<0$
and thus the absolute value of the integrand decays rapidly if we
depart from the real $s$-axis in the lower complex half-plane.
Imposing the even stronger constraint $h\ll\abs{\Im(\tilde{s})}\ll1$,  
the decay of the exponent dominates all the other $s$-dependences 
[$\gamma_1(s)$, $F_{01}(s)$, and $\Delta E(s)$] since their typical
(minimum \cite{well-behaved}) 
scale of variation is $\abs{\Im(\tilde{s})}\ll1$.  
In view of the complex continuation of Eq.~(\ref{Ecoeff2}), the same
applies to the amplitude $a_0(s)$.
As a result, the above integral~(\ref{Eformal}) will be exponentially 
suppressed $\sim\exp\{-\ord(\abs{\Im(\tilde{s})}/h)\}$ if
$h\ll\abs{\Im(\tilde{s})}\ll1$ holds, which (as one would expect)
implies a large  evolution time $T$ via the side condition $s(T)=1$. 

The general situation with varying $h(s)$ can be treated in complete
analogy -- the integral in Eq.~(\ref{Eformal}) is suppressed provided 
that the condition
\bea
\label{Ecriterion}
h(0)+h(1)\ll1
\,\wedge\,
\Re\left(i\int\limits_0^{\Re(\tilde{s})+i\Im(\tilde{s})/2}
\frac{ds}{h(s)}\right)
\gg1
\ea
holds for all singularities~$\tilde{s}$ 
(and saddle points etc.) in the lower complex half-plane 
(which determine the deformation of the integration contour).  
Together with
\bea
\label{Erun-time}
T=\int\limits_0^1\frac{ds}{\Delta E(s) h(s)}
\,,
\ea
this determines an upper bound for the necessary runtime $T$ of the
quantum adiabatic algorithm.

Note that the constraint $\dot s\ll\abs{\Im(\tilde{s})}\Delta E$ derived
from $h\ll\abs{\Im(\tilde{s})}$ is not necessarily equivalent
to $\dot s\ll\Delta E^2$, which one would naively deduce from
Eq.~(\ref{Eadiabatic_old}). 

\section{Evolution time}
%
The general criterion in Eq.~(\ref{Ecriterion}) can now be used to
estimate the necessary run-time via Eq.~(\ref{Erun-time}).
Typically, the inverse energy gap $1/\Delta E(s)$ is strongly peaked
(along the real axis) around $\Re(\tilde{s})$ with a width
\cite{well-behaved} of order $\abs{\Im(\tilde{s})}$.
Therefore, assuming $h(s)$ to be roughly constant across the peak and 
respecting $h\mid_{\rm peak}\ll\abs{\Im(\tilde{s})}$, yields the
following estimate of the integral in Eq.~(\ref{Erun-time}) 
\bea
\label{inverse}
T=\ord\left(\Delta E_{\rm min}^{-1}\right)
\,,
\ea
where $\Delta E_{\rm min}$ denotes the minimum energy gap.
Note that this estimate is only valid for one (or a few) relevant
excited state(s) -- multiple excited states will be discussed in
section \ref{Sextension}.

Intuitively, the same order of magnitude estimate for the evolution
time can also be derived from the local adiabatic condition 
(\ref{Eadiabatic_old}): 
Inverting this condition, we find the relationship
\bea
\label{inverse-adiabatic}
T=\frac1\varepsilon
\int\limits_0^1
ds\,
\frac{F_{01}(s)}{\Delta E^2(s)}
\,.
\eea
Assuming that $F_{01}(s)$ does not oscillate strongly, e.g., that the
ground state of $H(s)$ travels on a reasonably direct path from the
initial to the final state, we can make the following estimate 
\bea
T=
\frac{\ord(\Delta E_{\rm min}^{-1})}{\varepsilon}
\int\limits_0^1 ds\,
\frac{F_{01}(s)}{\Delta E(s)} 
\,.
\eea
Now we may exploit the advantage of the representation in
Eq.~(\ref{Eformal}), which is valid for general dynamics $s(t)$
corresponding to different functions $h(s)$ and hence for arbitrary 
evolution times $T$.
In the limit of very fast evolution $T\to 0$ (which implies 
$h \to \infty$), we have large excitations $a_1(T)=\ord(1)$ and thus 
the remaining integral in the above equation can be estimated via 
inserting this limit into Eq.~(\ref{Eformal}):
\bea
\label{fast}
\int\limits_0^1
ds\,
\frac{F_{01}(s)}{\Delta E(s)}
=\ord(1)
\,.
\ea
By comparing Eqs.~(\ref{fast}) and (\ref{inverse-adiabatic}), we again
obtain the estimate (\ref{inverse}). 
Note that the quantities $F_{01}(s)$ and $\Delta E(s)$ appearing in
the integrals in Eqs.~(\ref{inverse-adiabatic}-\ref{fast}) 
do not depend on the dynamics $s(t)$ which allows us to perform the
integration independently of $s(t)$. 

\begin{figure}
\includegraphics[width=8cm]{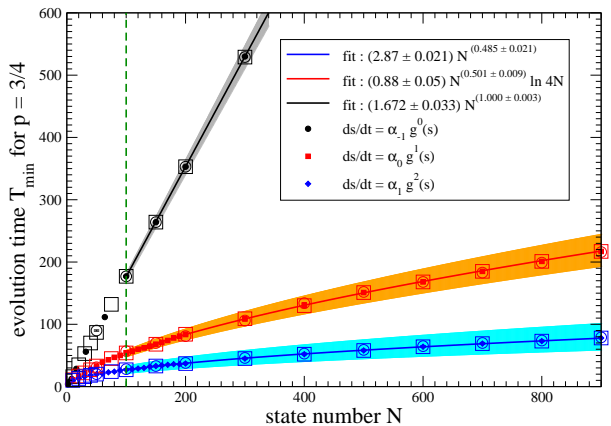}
\caption{\label{Fruntime}
[Color Online] Runtime scaling of the adiabatic 
Grover search for different interpolation functions $s(t)$ and a
target fidelity of $3/4$. 
Solid lines represent fits to full symbol data for $N \ge 100$ and
shaded regions correspond to fit uncertainties (99\% confidence level).
These uncertainties arise from the finite resolution when determining the
necessary runtime.
Hollow circles represent calculations with smoothed
$C^\infty$-interpolations (compare dotted lines in 
figure \ref{Ffullcomp} and section \ref{Sextension}), whereas 
hollow boxes correspond to the nonlinear interpolation example in
section \ref{Sextension}.
} 
\end{figure}

\subsection{Gap Structure}
%
Let us illustrate the above considerations by means of the rather
general ansatz for the behavior of the gap
\bea
\label{Eenergy_gap}
\Delta E(s)=\left[(s-s_{\rm min})^{2a}+\Delta E^b_{\rm min}\right]^{1/b}
\,,
\ea
with the minimal gap $0<\Delta E_{\rm min}\ll1$ at $s_{\rm min}\in(0,1)$,
$b>0$, and $a\in{\mathbb N}_+$.
An avoided level crossing in an effectively two-dimensional subspace 
corresponds to $2a=b=2$.
This is the typical situation if the commutator of the initial and the
final Hamiltonian $[H_{\rm i},H_{\rm f}]$ is small, since, in this case,
the two operators can almost be diagonalized independently and thus the 
energy levels are are nearly straight lines 
except at the avoided level crossing(s), where $[H_{\rm i},H_{\rm f}]$
becomes important.
In the continuum limit, such an (Landau-Zener type) avoided level
crossing corresponds to a second-order quantum phase transition.
The finite-size analogue of a third-order phase transition corresponds to 
$a=b$ (and accordingly for even higher orders), which may occur if 
$[H_{\rm i},H_{\rm f}]$ is not small or if the interpolation is not
linear, i.e., $H(s)\neq[1-s] H_{\rm i} + s H_{\rm f}$.

The inverse gap $1/\Delta E(s)$ has singularities around $s_{\rm min}$
at $\Im(\tilde{s})=\ord(\Delta E_{\rm min}^{b/2a})$,  
compare Fig.~\ref{Fintpath}.
The total running time $T$ for different choices of 
${h(s)=\alpha_d \Delta E^d(s)}$ 
satisfying the criterion~(\ref{Ecriterion})
can be obtained from Eq.~(\ref{Erun-time}).
Here, the exponent $d$ determines the scaling of the interpolation
dynamics, whereas the coefficient $\alpha_d$ is adapted such that
$s(T)=1$, cf.~Eqs.~(\ref{Eclassify}) and (\ref{Erun-time}).

For \mbox{$2a(d+1)/b>1$} one easily shows that 
\mbox{$1/\alpha_d=\ord(T\Delta E_{\rm min}^{d+1-b/2a})$} satisfies the 
criterion~(\ref{Ecriterion}) with the evolution time obeying 
\mbox{$T=\ord(\Delta E_{\rm min}^{-1})$}.
If $d$ is smaller, the necessary evolution time will be larger. 
In Table~\ref{Tscaling}, the scaling of the run-time 
(for two examples of the gap structure)
is derived for three cases:
\begin{itemize}
\item[a)] constant velocity $\dot s=\alpha_{-1}$, i.e., $d=-1$,
\item[b)] constant function $h(s)=\alpha_0$, i.e., $d=0$, and
\item[c)] the local adiabatic evolution with 
$h(s) = \alpha_1 \Delta E(s)$, i.e., $d=+1$, 
investigated in~\cite{roland2002}.
\end{itemize}

\begin{table}[h]
\begin{tabular}{|c|c|c|}
\hline
$\Delta E(s)=$
&
$\sqrt{(s-1/2)^2+\Delta E^2_{\rm min}}$ 
& 
$\sqrt{(s-1/2)^4+\Delta E^2_{\rm min}}$ 
\\
\hline
$d=-1$
&
$\Delta E^{-2}_{\rm min}$
&
$\Delta E^{-3/2}_{\rm min}$
\\
$d=0$
& 
$\Delta E^{-1}_{\rm min}\ln\Delta E^{-2}_{\rm min}$
&
$\Delta E^{-1}_{\rm min}$
\\
$d \ge 1$
&
$\Delta E^{-1}_{\rm min}$
&
$\Delta E^{-1}_{\rm min}$
\\
\hline
\end{tabular}
\caption{\label{Tscaling}
Scaling of the runtime $T$ necessary to obtain a fixed
fidelity for different gap structures (top row) and varying
interpolation velocities (first column).
The best improvement possible scales as the inverse of the minimum
gap $\Delta E^{-1}_{\rm min}$.}
\end{table}

\subsection{Grover's Algorithm}\label{Grover}
%
In the frequently studied adiabatic realization of Grover's algorithm
(see, e.g., \cite{childs2002,das2003,roland2002})
the initial Hamiltonian reads 
\mbox{$H_{\rm i}=\f{1}-\ket{{\rm in}}\bra{{\rm in}}$}
with the initial superposition state 
\mbox{$\ket{{\rm in}}=\sum_{x=0}^{N-1}\ket{x}/\sqrt{N}$},
and the final Hamiltonian is given by
\mbox{$H_{\rm f}=\f{1}-\ket{w}\bra{w}$}, 
where $\ket{w}$ denotes the marked state.
In this case, the commutator is very small
$[H_{\rm i},H_{\rm f}]=
(\ket{{\rm in}}\bra{w}-\ket{w}\bra{{\rm in}})/\sqrt{N}$
and one obtains for the time-dependent gap \cite{roland2002}
\bea
\label{Egap}
\Delta E(s) 
&=& 
\sqrt{1 - 4\left(1-\frac{1}{N}\right)s(1-s)}
\nn
&\approx&
\sqrt{4\left(s-\frac12\right)^2+\frac1N}
\,.
\eea
Comparing with Eq.~(\ref{Eenergy_gap}), we identify 
\mbox{$\Delta E_{\rm min}\approx 1/\sqrt{N}$} and $2a=b=2$
(the pre-factor does not affect the scaling behavior).
Consequently, our analytical estimate implies 
$T=\ord(N)$ for $d=-1$, 
$T=\ord(\sqrt{N}\ln 4 N)$ for $d=0$, and
$T=\ord(\sqrt{N})$ for $d>0$.

\begin{figure}
\includegraphics[width=8cm]{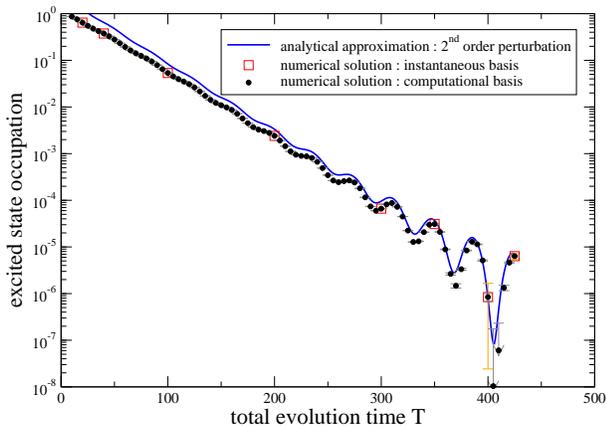}
\caption{\label{exp-decay}[Color Online]
Final error probability $|a_1(T)^2|$ as a function of run-time $T$ for 
Grover's algorithm with $N=100$ and $h=\rm const$.
The oscillations stem from the time-dependence of $a_0$ in 
Eq.~(\ref{Eformal}).
The solid (blue) line represents the second-order perturbative
solution of Eq.~(\ref{Eformal}).} 
\end{figure}

\begin{figure}
\includegraphics[width=8cm]{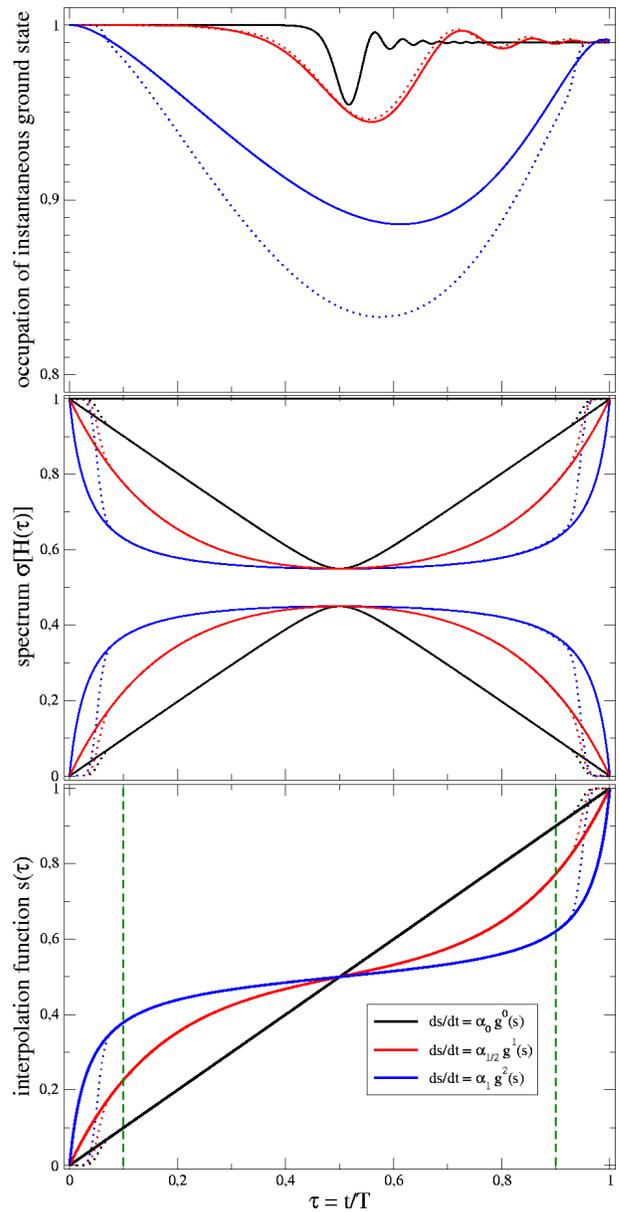}
\caption{\label{Ffullcomp}[Color Online]
Evolution of the interpolation function $s(t)$ ({\bf bottom panel}),
the spectrum $\sigma[s(t)]$ ({\bf middle panel}),
and the occupation of the instantaneous ground state ({\bf top panel})
versus the rescaled time $\tau=t/T$ for an adiabatic Grover search
problem with $N=100$ states. 
For each interpolation (different line styles), 
$T$ was adapted to reach 99\% of final fidelity. 
Thin dotted lines represent $C^\infty$-interpolations
smoothed with a test function.}  
\end{figure}

We have solved the Schr\"odinger equation numerically by using a 
fourth order Runge-Kutta integration scheme with an adaptive step-size 
\cite{press1994}. 
By restarting the code with different $T$ until agreement with desired
fidelity was sufficient, we could confirm these runtime scaling
predictions numerically, see Fig.~\ref{Fruntime}.  
The dependence of the final error on the run-time $T$ for fixed
$N=100$ and constant $h$ is depicted in Fig.~\ref{exp-decay}, where
the exponential decay becomes evident.
The evolution of the instantaneous ground-state occupation is plotted
in Fig.~\ref{Ffullcomp} for the three different dynamics.

\section{Further generalizations}\label{Sextension}

\subsection{Adiabatic Switching}
%
From an experimental point of view, the time-dependence of the
Hamiltonian will most certainly vanish asymptotically 
$\dot H(t<0)=\dot H(t>T)=0$ or at least be negligible -- 
which automatically implies $h(0)=h(1)=0$.
Furthermore, realistic Hamiltonians should be described by 
$C^\infty$-interpolations ({\em Natura non facit saltus}).

By using a $C^\infty$-test function which was matched at 
$t_1=0.1 T$ and $t_2=0.9 T$ to the usual dynamics $s(t)$
(compare dotted lines in figure \ref{Ffullcomp} bottom panel), 
we have implemented an interpolation scheme with 
such an adiabatic switching on and off 
$\dot{s}(0) = \dot{s}(T) = 0$.
For the investigated adiabatic implementation of the Grover search
routine, this scheme does not affect the final result considerably.
The reason for this robustness lies in the fact that the matrix
element $F_{nm}$ is peaked around $s=1/2$ and $h(0)$ as well as $h(1)$
are small enough already without the adiabatic switching on and off.
Therefore, one can expect the dominant non-adiabatic corrections to
arise from the behavior around the minimum gap, which was unaffected
by the test function.
This is also confirmed by the scaling of the runtime versus the system
size, compare the hollow circle symbols in figure \ref{Fruntime},
which is basically unchanged.

\begin{figure}
\vspace{0.5cm}
\includegraphics[width=8cm]{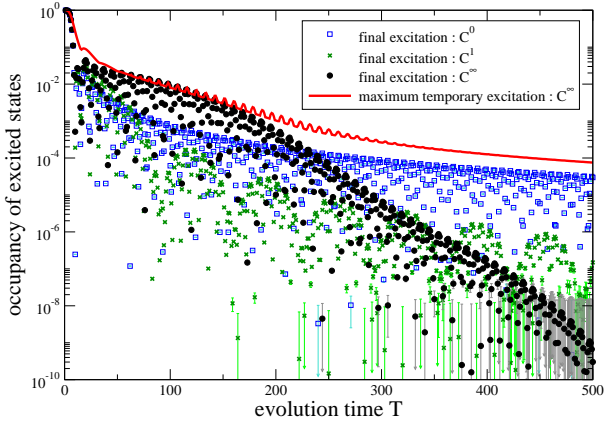}
\caption{\label{Ferror_example}[Color Online]
Evolution of the final and the maximum intermediate (red line)
excitations with the runtime $T$ for the
example~(\ref{Eexample_igs}). 
The exponential falloff in the final excitations is only visible, if a  
smooth $C^\infty$-interpolation (black circles) is used, whereas the
scaling of the intermediary excitations (red line) is always
polynomial. 
The suppression of the final error for $C^0$ or $C^1$-interpolations
(blue squares and green crosses) is also merely polynomial.}
\end{figure}

However, the situation is completely different for the example
considered in section \ref{Degeneracy} below.
There, the exponential suppression of the final error as a
function of the run-time requires a smooth $C^\infty$-interpolation --
with other dynamics such as $C^0$ (just continuous) or $C^1$ 
(differentiable once), the final error is merely polynomially small,
cf.~figure~\ref{Ferror_example}. 

\subsection{Nonlinear Interpolation}
%
Although we have chosen a linear interpolation 
scheme~(\ref{Ehamiltonevolution}) in order to satisfy the trace 
constraint~(\ref{Etraceconst}), the presented analysis can be
generalized easily to more general non-linear interpolations.
[Note that, {\em linear} refers to the straight connection line
between initial and final Hamiltonian in equation
(\ref{Ehamiltonevolution}) and should not be confused with the
different velocities $s(t)$ at which this line is traversed.]
The argumentation based on the analytic continuation works in the same
way provided that the functional dependence 
$H_{\rm nl}(s)=f(H_{\rm i},H_{\rm f},s)$ does not involve extremely
large or small numbers.

As an illustrative example, we consider the Grover search with the
same initial and final Hamiltonians but a quadratic interpolation
scheme 
\bea
H_{\rm nl}(s) 
&=&
[(1-s) H_{\rm i} + s H_{\rm f}]^2 + s(1-s)\,\frac{2N - 2}{N^2}\,\f{1}
\nn
&=& 
(1-s)^2 H_{\rm i} + s^2 H_{\rm f}
\nn
&&+ 
s(1-s)\left[\{H_{\rm i}, H_{\rm f}\}
+ \frac{2N - 2}{N^2}\,\f{1}\right]\,,
\eea
where $\{\cdot,\cdot\}$ denotes the anti-commutator.
The identity operator $\f{1}$ has been added in order to ensure 
${\rm Tr}\{H_{\rm nl}\}=N-1$, cf.~equation~(\ref{Etraceconst}).
Although the spectrum of this non-linear interpolation is slightly
distorted compared to the linear one, the fundamental gap is the same 
as in equation (\ref{Egap}), and hence same interpolation functions
$s(t)$, applied to the above Hamiltonian, should reproduce the
aforementioned scaling predictions.
This is confirmed by the numerical analysis of the scaling behavior
-- the results of the non-linear interpolation are basically
indistinguishable from those of the previous example 
(linear interpolation), compare the hollow box symbols in 
figure~\ref{Fruntime}. 

\subsection{Degeneracy}\label{Degeneracy}
%
So far, we have restricted our considerations to the instantaneous
ground state and a single first excited state. 
Let us now consider a very simple example (see also \cite{farhi0512159}) 
in which there is still a unique ground state, but many degenerate
first excited states: 
In terms of single-qubit Pauli matrices $\sigma_x$ and $\sigma_z$, the
$M$-qubit Hamiltonian reads 
\bea
\label{Eexample_igs}
H(s)=\frac{1}{2}\sum_{j=1}^M
\left[\f{1} - s\sigma_z - (1-s)\sigma_x\right]^{(j)}
\,,
\eea
where we have used a linear interpolation~(\ref{Ehamiltonevolution})
for simplicity. 
In this example, the Hamiltonian can be decomposed completely
into independent and equal single-qubit contributions and hence the
time-evolution operator factorizes, i.e., it is sufficient to solve
the dynamics of a single qubit.
Furthermore, the Hamiltonian is invariant under any permutation of the
qubits. 
The instantaneous ground states for all values of $s$ are symmetric
under this permutation group and hence unique, but the first excited
states are not -- leading to a $M$-fold degeneracy 
(i.e., there are $M$ equivalent first excited states).
Hence, the fundamental gap between the ground state and each one of
these first excited states is the same as for one qubit and thus 
independent of the number of qubits 
$\Delta E(s)=\sqrt{1-2 s(1-s)}$. 

In some sense, this simple example represents a limiting case opposite
to Grover's algorithm:
The energy gap $\Delta E(s)$ and the matrix elements $F_{nm}(s)$ do  
not scale with the number $M$ of qubits and the $F_{nm}$ are neither
small initially nor finally. 
Instead, the scaling with system size manifests itself in the $M$-fold
degeneracy of the first excited states.
As a result of the $M$-independent gap structure, the adiabatic
switching is crucial for achieving the exponential suppression of the
final error.
Figure~\ref{Ferror_example} displays the final error probabilities for
a smooth $C^\infty$-interpolation and for $C^0$ and
$C^1$-interpolations for comparison.
These numerical simulations confirm that the falloff is exponential in
the $C^\infty$-case but merely polynomial for $C^0$ and $C^1$.

\begin{figure}
\includegraphics[width=7.5cm]{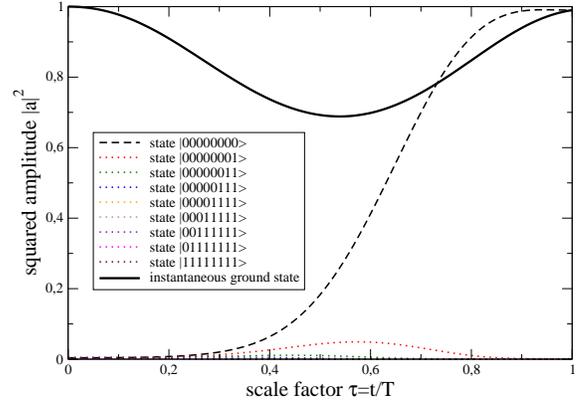}
\caption{\label{Fexample_igs}[Color Online]
Occupation of the instantaneous ground state and some selected
computational basis states for the Hamiltonian in~(\ref{Eexample_igs})
for an $M=8$ qubit system. 
Temporarily, the system leaves the instantaneous ground state, but the
runtime $T$ has been adjusted such that the final fidelity is 99\%.
}
\end{figure}
 
Another interesting point of this simple example is the difference
between the intermediate and the final occupation of the ground
state, see figures~\ref{Fexample_igs} and \ref{Ferror_example}.
According to the first-order result in Eq.~(\ref{Efirst_order}) and
the aforementioned factorization of the time-evolution operator, the 
intermediate excitation probability scales as 
\bea
\label{first-order-error}
p_{\rm int}
=
\sum\limits_{m>0}|a_m|^2
=
\ord\left(\frac{M}{T^2\Delta E^4}\right)
=
\ord\left(\frac{M}{T^2}\right)
\,,
\ea
since the gap $\Delta E$ is independent of $M$.
On the other hand, the final error probability 
(assuming a $C^\infty$-interpolation) is exponentially suppressed 
\bea
\label{final-error}
p_{\rm fin}
=\ord\left(M\exp\left\{-T\Delta E\right\}\right) 
=\ord\left(M\exp\left\{-T\right\}\right) 
\,,
\ea
and hence the two error probabilities can be vastly different 
$p_{\rm int} \gg p_{\rm fin}$, cf.~figure~\ref{Fexample_igs}. 
In fact, by increasing the number of qubits, the occupancy of the
instantaneous ground state can be made arbitrarily small.
Moreover, the run-time condition derived from the 
first-order result in Eqs.~(\ref{Efirst_order}) and
(\ref{first-order-error}) 
\bea
\label{first-order-run-time}
T_0=\ord(\sqrt{M})
\,,
\ea
yields a scaling which is far too pessimistic compared with the
correct final error probability assuming a $C^\infty$-interpolation
\bea
\label{final-run-time}
T_\infty=\ord(\ln M)
\,.
\ea
Note that non-smooth interpolations (e.g., $C^0$ or $C^1$) would also
yield a polynomial scaling $T=\ord(M^x)$ similar to 
Eq.~(\ref{first-order-run-time}).
On the other hand, the scaling behavior in Eqs.~(\ref{final-error})
and (\ref{final-run-time}) is just what one would obtain by immersing
the system in Eq.~(\ref{Eexample_igs}) into a zero-temperature
environment and letting it decay towards its ground state.
Therefore, using non-smooth interpolations (e.g., $C^0$ or $C^1$) or
naively demanding the first-order estimate in
Eq.~(\ref{Efirst_order}), the adiabatic algorithm would be even slower
than this simple decay mechanism.   

\section{Summary}
%
The instantaneous occupation of the first excited state {\em during} 
the adiabatic evolution in Eqs.~(\ref{Efirst_order}) and 
(\ref{Eadiabatic_old}) does not provide a good error estimate.
Instead, a better estimate is given by the remaining real excitations
{\em after} the dynamics. 
For the example plotted in Fig.~\ref{Ffullcomp}, the instantaneous
excitation probability exceeds 10\% at intermediate times -- whereas
the final value is 1\%.
This is even more drastic for the example in section~\ref{Degeneracy},
see figure~\ref{Fexample_igs}, where the two values and hence the
inferred run-times can differ by orders of magnitude.

Moreover, the final error can be made extremely -- 
in fact, with $h(0)+h(1)\lll1$, exponentially -- small
\bea
\label{summary-eq}
a_1(T)=\ord\left(h(0)+h(1)+
\exp\left\{-\frac{\abs{\Im(\tilde{s})}}{h(s_{\rm min})}\right\}\right)
\,,
\ea
cf.~Fig.~\ref{exp-decay}.
For the Grover example, the last term was dominant, whereas in the
general case the smallness of the first two terms can be ensured by
using smoothed $C^\infty$-interpolations, i.e., adiabatic switching --
which is a more realistic ansatz anyway. 

Based on general arguments, the optimal run-time 
(in the absence of degeneracy, cf.~section~\ref{Degeneracy})
scales as \mbox{$T=\ord(\Delta E^{-1}_{\rm min})$} contrary to what
one might expect from the Landau-Zener \cite{landau-zener} formula 
(with \mbox{$T\propto\Delta E^{-2}_{\rm min}$}).
In view of the fact that the minimum energy gap $\Delta E_{\rm min}$
is a measure of the coupling between the known initial state and the
unknown final state, this result is very natural.

For the Grover algorithm, it is known that the $\sqrt{N}$-scaling is
optimal \cite{roland2002}.
This optimal scaling \mbox{$T=\ord(\Delta E^{-1}_{\rm min})$} can
already be achieved with interpolation functions $s(t)$ which vary
less strongly (e.g., $d=0$) than demanded by locally \cite{roland2002}
adiabatic evolution ($d=1$) -- and hence should be easier to
realize experimentally.  

Unfortunately, a constant velocity with $d=-1$ does not produce the 
optimal result in general.
The Grover example has the advantage that the spectrum can be
determined analytically, which is for example not the case for the
more involved satisfiability problems \cite{farhi2000}.
Therefore, some knowledge of the spectral properties $\Delta E(s)$ is
necessary for achieving the optimal result 
\mbox{$T=\ord(\Delta E^{-1}_{\rm min})$} also in the general case of
adiabatic quantum computing.
For systems with an analytically unknown gap structure, some knowledge
about the spectrum can be obtained by extrapolating the scaling
behavior of small systems.

A related interesting point is the impact of the gap structure 
(corresponding to $2^{\rm nd}$ or $3^{\rm rd}$ order transition etc.)
in Eq.~(\ref{Eenergy_gap}).
The derived constraint for the velocity at the transition
$\dot s\ll\abs{\Im(\tilde{s})}\Delta E$ is only for $2^{\rm nd}$-order
transitions equivalent to $\dot s\ll\Delta E^2$, which one
would naively deduce from  Eq.~(\ref{Eadiabatic_old}).


Note that the improvement 
\mbox{$T=\ord(\Delta E^{-1}_{\rm min})$}
compared with the conventional runtime estimate 
\mbox{$T=\ord(\Delta E^{-2}_{\rm min})$}
is merely polynomial (same complexity class). 
Though this is not as impressive as an exponential speedup, in
practice a polynomial improvement may be useful.
For time-dependent Hamiltonians where the inverse of the minimum gap
scales exponentially with the size of the problem, we would still
expect an exponential scaling of the runtime $T$ required to reach a
fixed fidelity (as in section~\ref{Grover}).
On the other hand, the exponential suppression of the final error in  
Eq.~(\ref{summary-eq}) may become important in certain cases such as
in the presence of degeneracy and may well yield an exponential
speedup in comparison with the conventional estimate, see
section~\ref{Degeneracy}. 

In some sense, the two examples in sections~\ref{Grover}
and~\ref{Degeneracy} represent two simple extremal examples for
adiabatic quantum computing regarding the scaling of the gap and the
degeneracy. 
For more complicated situations such as satisfiability problems
\cite{farhi2000}, both properties have to be taken into account
simultaneously. 

\section*{Acknowledgments}
%
R.~S.~acknowledges fruitful discussions during the workshop 
"Low dimensional Systems in Quantum Optics"  
at the CIC in Cuernavaca (Mexico), 
which was supported by the Humboldt foundation. 
G.~S.~acknowledges fruitful discussions with M.~Tiersch.
This work was supported by the Emmy Noether Programme of the German
Research Foundation (DFG) under grant No.~SCHU~1557/1-1/2.

\section*{Note added}
%
Recently, two of the main results of this article, i.e., the optimal
run-time scaling \mbox{$T=\ord(\Delta E^{-1}_{\rm min})$} and the
faster-than-polynomial decrease of the final error $a_1(T)$, have been
demonstrated rigorously for a class of Hamiltonians using methods of
spectral analysis \cite{jansen0603175}. 



\begin{thebibliography}{499}
\bibitem{well-behaved}
Since the functions $\gamma_1(s)$, $F_{01}(s)$, and $\Delta E(s)$  
are supposed to be well-behaved near the real $s$-axis, there are no
small (or large) numbers in the problem apart from those generated by
the minimum of the gap $\Delta E(s)$.
%
Thus the significant changes of the eigenvectors are also localized 
around this minimum. 
%
In the complex plane, this minimum along the real axis becomes a
saddle point.
%
For analytic functions, the characteristic length scale of variation 
must be the same along the real axis and into the complex plane
(of order $\abs{\Im(\tilde{s})}\ll1$) and is determined by the lowest 
non-trivial Taylor coefficient at that point.
\end{thebibliography}
\end{document}